\documentclass[prl,aps,twocolumn,superscriptaddress,floats,showpacs,amsmath,amssymb]{revtex4}

\usepackage{graphicx}
\usepackage{units}

\def \be{\begin{equation}}
\def \ee{\end{equation}}

\newcommand{\saclay}{%
           Nanoelectronics group, Service de Physique de l'Etat Condens{\'e}, 
           CEA Saclay F-91191 Gif-sur-Yvette Cedex, France}

 \newcommand{\thales}{%
           Unit\'e Mixte de Physique CNRS-Thales, Route d\'epartementale 128,
91767 Palaiseau Cedex and Universit\'e Paris-Sud 91405, Orsay, France}

\begin{document}

\title{Spin torque and waviness in magnetic multilayers: 
a bridge between Valet-Fert theory and quantum approaches}

\author{Valentin S. Rychkov}
\affiliation{\saclay}
\author{Simone Borlenghi}
\affiliation{\saclay}
\author{Henri Jaffres}
\affiliation{\thales}
\author{Albert Fert}
\affiliation{\thales}
\author{Xavier Waintal}
\affiliation{\saclay}

\date{\today}

\begin{abstract}
We develop a simple theoretical framework for transport in magnetic 
multilayers, based on Landauer-Buttiker scattering formalism and 
Random Matrix Theory. A simple transformation allows one to go from the scattering point
of view to theories expressed in terms of local currents and electrochemical potential.
In particular, our theory can be mapped onto the well established classical Valet Fert theory for
collinear systems. For non collinear systems, in the absence of spin-flip scattering, 
our theory can be mapped onto the generalized circuit theory. 
We apply our theory to the angular dependance of
spin accumulation and spin torque in non-collinear spin valves.  
\end{abstract}


\maketitle

The discovery of Giant Magnetoresistance (GMR)~\cite{GMR_Fert_Grun} and its subsequent counterpart, the
current induced spin torque~\cite{Spin_torque_Slon}, was at the origin of a new field 
which aims at controlling the magnetization dynamics of small
metallic devices through standard electronics.
The theory of transport in those systems is now well developed and includes a
number of approaches that range from classical Valet Fert (VF) theory in the 
diffusive regime~\cite{ValetFert_PRB} , Boltzmann equation~\cite{Boltzman} to quantum approaches (original~\cite{Circuit_Theory} and generalized~\cite{General_Circuit_Theory}) circuit theory, random matrix theory (RMT)
for the scattering matrix ~\cite{Waintal_RMT}, and ab-initio based models~\cite{Ab_initio_ballistic, Ab_initio_TB}.
Many connections exist between these different approaches.  One popular route~\cite{BrataasBauerKelly} starts from the Keldysh Green's function 
formalism. In the quasi-classical approximation it yields the Boltzmann equation and the (VF) diffusive equation. A different 
strategy where local equilibrium is assumed only in certain points leads to circuit theory.
An alternative route is the Landauer-B\"uttiker formalism which expresses the problem of
transport inside a quantum conductor as a scattering problem where electrons undergo various
transmission/reflection events as they go through the system. This approach is well suited for 
coherent system and is equivalent to the Keldysh approach. However, the classical
concepts of chemical potential or local equilibrium do not arise naturally in the scattering
approach so that classical intuitions do not easily transfer into its language.

In this letter we take the scattering formalism as our starting point and develop a theory which fully captures 
VF and (generalized) circuit theory. Our theory (here after referred as C-RMT for Continuous Random Matrix Theory) 
can be tabulated by the same set of (experimentally accessible)
parameters as VF~\cite{BassPratt}. 
On the other hand, it properly includes Sharvin resistance and allows for non collinear
and even a one-dimensional texture of magnetization (i.e. domain walls). 
We apply C-RMT to the discussion of the angular dependance of spin torque 
(``waviness''~\cite{Barnas_wavy,Boulle_wavy}) in asymmetric spin valves.

\begin{figure}
\includegraphics[keepaspectratio,width=0.8\linewidth]{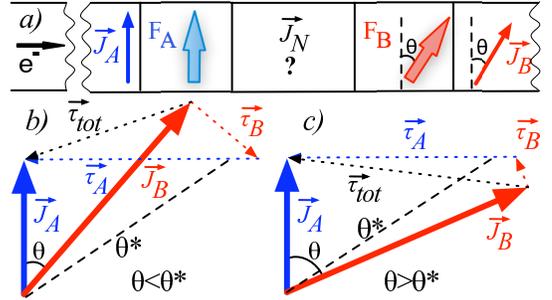}
\caption{\label{fig:cartoon} (a) Schematic of a spin valve with two ferromagnetic layer $F_A$ and $F_B$ whose magnetization makes an angle $\theta$. (b) and 
(c): geometric construction of the spin torque, see text. $\vec J_A$, $\vec J_N$
and $\vec J_B$ are the spin currents along the valve while $\vec \tau_A$
and $\vec \tau_B$ are the torque on the two magnetic layers.}
\end{figure}

{\it Two quantum conductors in series.} Before introducing C-RMT, we start with an instructive example of transport in a non-magnetic metal. C-RMT is a mere extension of the concepts developed below to properly take into account the electronic spins. Classically, the electronic transport in a 
system $A$ is described by its resistance ${\cal R}_A$. When two systems $A$ and $B$ 
are put in series,
the total resistance is described by Ohm law ${\cal R}_{AB}={\cal R}_A+{\cal R}_B$.
In a coherent quantum system, the transport properties can be described by the amplitudes for
reflection and transmission, and the conductor is characterized by its
scattering matrix $S_A$,
\be
S_A=
\left(\begin{array}{cc}  r'_A & t_A \\ t'_A & r_A \end{array}\right)
\ee
where $r_A$ and $t_A$  ($t'_A$ and $r'_A$) describe the reflection/transmission amplitudes
to the right (left) of the sample. 
When putting two such conductors in series, one has to add the amplitudes for all possible 
processes. For instance $t_{AB}= t_B t_A + t_B (r_A r'_B) t_A + t_B (r_A r'_B)(r_A r'_B) t_A +\dots$
includes the direct process where the electron is transmitted by $A$ and then by $B$ ($t_B t_A$),
the process where the electron is transmitted by $A$, then reflected by $B$ and $A$, and
then transmitted by $B$ ($ t_B r_A r'_B t_A$) and so on. The previous geometrical series
can be resumed into,
\begin{eqnarray}
\label{coh:add}
t_{AB} &=&t_B[1- r_A r'_B]^{-1} t_A \\
\label{coh:add_r}
r_{AB} &=&r_B + t_B[1- r_A r'_B]^{-1}r_A t'_B. 
\end{eqnarray}
The conductance of such a system is given by the Landauer formula $g_{AB}= (e^2/h) T_{AB}$
where $T_{AB}=|t_{AB}|^2$ is the probability for an electron to be transmitted. Let us now
suppose that many channels $N_{\rm ch}$ of propagation are present in the system, so that the
transmission probability has to be averaged over those. The product  $r'_B r_A=|r'_B r_A| z$ contains a phase $z=e^{i\phi}$ which is almost random from channel to channel, so that with very good approximation the total transmission is given
by its average over $z$. Straightforward integration gives, 
\be
\label{addition}
T_{AB}= T_B (1- R_A R'_B)^{-1} T_A
\ee
where $R_A=|r_A|^2=1-T_A$ is the reflection probability. Eq.(\ref{addition})
can be recast into $1/T_{AB}=1/T_A +1/T_B -1$ which leads to 
$ {\cal R}_{AB}={\cal R}_A+{\cal R}_B - {\cal R}_{sh} $
where the Sharvin resistance is given by ${\cal R}_{sh}= h/(e^2 N_{\rm ch})$.
The above equations call for a number of comments. (i) Addition law for resistances is very close
to Ohm law except for the presence of the Sharvin (or contact) resistance, but for classical conductors,
${\cal R}_{A(B)}\gg {\cal R}_{sh}$ one recovers Ohm law. (ii) Eq.(\ref{addition}) has a
structure completely similar to Eq.(\ref{coh:add}). In fact, it can be obtained directly in
a way similar to Eq.(\ref{coh:add}) by adding not the amplitudes but the probabilities for
the various processes. (iv)  We note
that Ohm law is derived here using purely quantum mechanical concepts for fully coherent
conductors. In particular there is no need for a well defined chemical potential 
(local equilibrium) in between the two conductors $A$ and $B$ even though everything 
happens as if such a chemical potential was defined. 

{\it RMT.} We now proceed with the extension of the above ideas to magnetic systems where
the $S$ matrix now includes a spin grading.
Our starting point is an extension of RMT that was introduced in Ref~\cite{Waintal_RMT}. The theory
has a structure almost identical to the scattering approach except that a conductor (or a
part of it) is now described by $4\times 4$ ``hat'' matrices $\hat{r}$, $\hat{r}'$,
$\hat{t}$ and $\hat{t}'$ defined in term of the ($4 N_{\rm ch}\times 4 N_{\rm ch}$) $S$
matrix,
\be
\hat{r}=\frac{1}{N_{\rm ch}} {\rm Tr}_{N_{\rm ch}}
\left(
\begin{array}{cccc} r_{\uparrow \uparrow} r_{\uparrow \uparrow}^{\dagger} &
r_{\uparrow \uparrow}r_{\uparrow \downarrow}^{\dagger} &
 		    r_{\uparrow \downarrow}r_{\uparrow \uparrow}^{\dagger} &
r_{\uparrow \downarrow}r_{\uparrow \downarrow}^{\dagger} \\
 		    r_{\uparrow \uparrow}r_{\downarrow \uparrow}^{\dagger} &
r_{\uparrow \uparrow}r_{\downarrow \downarrow}^{\dagger} &
 		    r_{\uparrow \downarrow}r_{\downarrow \uparrow}^{\dagger} &
r_{\uparrow \downarrow}r_{\downarrow \downarrow}^{\dagger} \\
                    r_{\downarrow \uparrow}r_{\uparrow \uparrow}^{\dagger}
&  r_{\downarrow \uparrow}r_{\uparrow \downarrow}^{\dagger} &
 		    r_{\downarrow \downarrow}r_{\uparrow \uparrow}^{\dagger} &
r_{\downarrow \downarrow}r_{\uparrow \downarrow}^{\dagger} \\
                    r_{\downarrow \uparrow}r_{\downarrow
\uparrow}^{\dagger} &  r_{\downarrow \uparrow}r_{\downarrow
\downarrow}^{\dagger} &
 		    r_{\downarrow \downarrow}r_{\downarrow \uparrow}^{\dagger} &
r_{\downarrow \downarrow}r_{\downarrow \downarrow}^{\dagger}
\end{array} \right).
\ee
 with similar definitions for $\hat{r}'$, $\hat{t}$ and $\hat{t}'$. The main result of
Ref.~\cite{Waintal_RMT} is that the addition law for the "hat" matrices is exactly given by
Eq.(\ref{coh:add}) and (\ref{coh:add_r}) except that now the transmission and reflection matrices are to be replaced by their "hat" counterparts. The conductance is given by
$
g= \left( \hat{t}_{11} +\hat{t}_{14}+\hat{t}_{41}+\hat{t}_{44} \right)/{{\cal R}_{\rm sh}} 
=(1/{{\cal R}_{\rm sh}})\sum_{\sigma\sigma'} T_{\sigma\sigma'}$, where $T_{\sigma\sigma'}=(1/N_{\rm ch}) {\rm Tr} t_{\sigma\sigma'} t^\dagger_{\sigma\sigma'}$
is the probability for an electron with spin $\sigma'$ to be transmitted with 
a spin $\sigma$. The "hat'' matrices for the interfaces between
two metals (say copper and cobalt) can be obtained from ab-initio calculation~\cite{Ab_initio_Stiles,Ab_initio_Kelly}, 
or tabulated from the experiments (see below). Once the ``hat'' matrices are known
for one direction of the magnetization, they can be rotated in arbitrary direction
using rotation matrices~\cite{Waintal_RMT}. 

{\it C-RMT.} To obtain the "hat'' matrices for
the bulk parts of the metallic layers, we need to proceed further. Assuming we know
the ``hat'' matrix $\hat S(L)$ for a system of size $L$, we can add an infinitely small
layer of size $\delta L$ and compute $\hat S(L+\delta L)$
from the knowledge of $\hat S(\delta L)$  and the addition law for "hat" matrices. 
The generic form for $\hat S(\delta L)$ is given by,
\be
\label{eq:thin}
\hat t = 1 - \Lambda^t \delta L  \ \ , \ \ \hat r = \Lambda^r \delta L
\ee
where the two matrices $\Lambda^t$ and $\Lambda^r$ entirely characterize the bulk
properties of the material. Expanding the addition law Eq.(\ref{coh:add})
in $\delta L$ provides,
\begin{eqnarray}
\label{drdl}
\partial\hat r/\partial L &=& \Lambda^r -\Lambda^t\hat r -\hat r\Lambda^t 
+\hat r\Lambda^r\hat r \\
\label{dtdl}
\partial\hat t/\partial L &=& -\Lambda^t\hat t + \hat r\Lambda^r t
\end{eqnarray}
 The various elements entering 
in the "hat" matrices (say $\hat t$) 
are vastly inequivalent. The main elements are the probabilities 
$T_{\sigma\sigma'}$. Secondly comes the element on the diagonal, the so called
(complex) mixed transmission 
$T_{\rm mx}=(1/N_{\rm ch}) {\rm Tr} t_{\uparrow \uparrow}t_{\downarrow \downarrow}^{\dagger}$
which measures how a spin transverse to the magnetic layer can be transmitted through the
system. These elements are usually believed to be rather small~\cite{Stiles_Mixing,Ab_initio_Kelly} 
in magnetic systems, but can play a role in non-collinear configurations nevertheless. Last,
the other elements involve
some coherence between spin-flip and non spin-flip processes and are likely to be even smaller.
In the basis parallel to a layer magnetization, they can be disregarded. We
parametrize a bulk layer by four parameters $\Gamma_\uparrow$, $\Gamma_\downarrow$ $\Gamma_{\rm sf}$ and $\Gamma_{\rm mx}$,
 \be
\label{eq:lambda}
\Lambda^t=
\left(\begin{array}{cccc}  
\Gamma_\uparrow +\Gamma_{\rm sf} & 0              &  0             & -\Gamma_{\rm sf} \\ 
           0                     & \Gamma_{\rm mx}&  0             & 0 \\
           0                     & 0              &\Gamma^*_{\rm mx} & 0 \\
        -\Gamma_{\rm sf}         & 0              & 0              & \Gamma_\downarrow+\Gamma_{\rm sf} 
\end{array}\right)
\ee
$l_\sigma=1/\Gamma_\sigma$ is the mean free path for spin $\sigma$. In a ferromagnet, $\Gamma_{\rm mx}= 1/l_\perp+i/l_L$
where $l_\perp$ is the penetration length of transverse spin current inside the magnet while $l_L$ is the
Larmor precession length. Those lengths, which are roughly equal, are the smallest characteristic lengths
with typical values smaller than $1$nm. In a normal metal, $\Gamma_\uparrow=\Gamma_\downarrow=\Gamma$ and
$ \Gamma_{\rm mx}=\Gamma+2\Gamma_{\rm sf}$ so that the ``hat'' matrices remain invariant upon arbitrary rotation of
the spin quantization axis. $\Lambda^r$ is given by the same parametrization as Eq.(\ref{eq:lambda}) 
with $\Gamma_{\rm sf}$ being replaced by $-\Gamma_{\rm sf}$ (in order to fulfill current conservation)
and neglecting $\Gamma_{\rm mx}$ for ferromagnets (as the mixing conductance is essentially of ballistic origin). 
This completes the formulation of the theory. Eq.(\ref{drdl}) and Eq.(\ref{dtdl})
can be integrated to obtain the "hat" matrices of the bulk parts. A given multilayer is then constructed 
by using the addition law Eq.(\ref{coh:add}) and (\ref{coh:add_r}) for the various bulk layers and 
the corresponding interfaces. 


{\it Link with Valet Fert theory.} Let us introduce the $4$-vector 
${\bold P}_\pm(x)=(P_{\pm,\uparrow}, P_{\pm,\rm mx},P_{\pm,\rm mx}^*, P_{\pm,\downarrow}$ where
$P_{+,\uparrow}(x)$ ($P_{-,\downarrow}$) is the probability to find a left (right) moving electron with spin
up (down) at point $x$. The addition law Eq.(\ref{coh:add}) and (\ref{coh:add_r}) (for "hat" matrices)
is equivalent to state that ${\bold P}_\pm(x_1)$ and ${\bold P}_\pm(x_2)$ on two sides of a conductor $A$
are related through its "hat" scattering matrix $\hat S_A$ as,
\be
\label{eq:master}
\left[\begin{array}{c}  {\bold P}_-(x_1) \\
                        {\bold P}_+(x_2) 
\end{array}\right]
= \hat S_A
\left[\begin{array}{c}  {\bold P}_+(x_1) \\
                        {\bold P}_-(x_2) 
\end{array}\right]
\ee
In the physical picture where one represents the scattering processes as random events with certain
 transmission or reflection probabilities, the above equation has a simple interpretation when one focuses 
on its first and fourth raw 
$P_{\pm,\uparrow}$ and $P_{\pm,\downarrow}$: it accounts for the conservation of probability in the scattering events, i.e. it is the Master equation of the underlying Brownian motion undertaken by the incident electrons.
Let us now introduce two new  $4$-vectors, ${\bold j}(x)$ and $\mu(x)$ defined as,
\begin{eqnarray}
\label{eq:j}
{\bold j}(x) &=&  [{\bold P}_+(x) - {\bold P}_-(x)]/(e{\cal R}_{\rm sh})\\
\label{eq:mu}
\mu(x) &=& [{\bold P}_+(x) + {\bold P}_-(x)]/2
\end{eqnarray}


(with $e<0$). Now, using the parametrization Eq.(\ref{eq:thin}) and Eq.(\ref{eq:lambda}) for infinitely thin layer, and writing Eq.(\ref{eq:master})  in term of ${\bold j}(x)$ and ${\mu}(x)$, we arrive at,
\begin{eqnarray}
\label{eq:vf}
j_\sigma &=& -1/(e\Gamma_\sigma{\cal R}_{\rm sh})\  \partial_x \mu_\sigma \\
\partial_x j_\sigma &=& 4 \Gamma_{\rm sf} / (e{\cal R}_{\rm sh})\  [\mu_{-\sigma} - \mu_\sigma]
\end{eqnarray}
which are precisely the VF equations~\cite{ValetFert_PRB}. Hence, for a collinear system, C-RMT simply reduces to
VF theory. In its original form however, VF theory does not account for the presence of Sharvin resistance.
Here the boundary conditions are given by the Landauer formula: the presence of a potential drop $eV$ between the reservoirs located at, say, $x=0$
and $x=L$ imposes $P_{+\sigma}(0)=eV$ and $P_{-\sigma}(L)=0$ which translates into,
\begin{eqnarray}
\label{eq:bc}
\mu_\sigma(0) +(e{\cal R}_{\rm sh}/2)\  j_\sigma(0)&=&eV \\
\mu_\sigma(L) - (e{\cal R}_{\rm sh}/2)\  j_\sigma(L)&=&0 
\end{eqnarray}
These mixed boundary conditions allows VF theory to properly include the Sharvin resistance of
the system and correspond to adding half of the Sharvin resistance on both extremities of the sample.

{\it Link with generalized circuit  theory.} Let us now consider Eq.(\ref{eq:master}) for a conductor
whose transmission and reflection matrices are purely diagonal, i.e. without any spin-flip scattering.
Let us further suppose that $R_{\rm mx}$ might be non zero but $T_{\rm mx}=0$. Then, Eq.(\ref{eq:master}) 
takes the form
\begin{eqnarray}
\label{eq:circuit1}
j_\sigma(x_1)=j_\sigma(x_2)&=&\frac{1}{{e\cal R}_{\rm sh}}
\frac{T_\sigma}{1-T_\sigma} [\mu_\sigma(x_1)-\mu_\sigma(x_2)]\\
\label{eq:circuit2}
j_{\rm mx}(x_{1,2}) &=& \pm \frac{2}{{e\cal R}_{\rm sh}}
\frac{1-R_{\rm mx}^{1,2}}{1+R_{\rm mx}^{1,2}} \mu_{\rm mx}(x_{1,2})
\end{eqnarray}
where $T_\sigma\equiv T_{\downarrow\sigma} + T_{\uparrow\sigma}$ is the total transmission of an incident electron
with spin $\sigma$ and $R_{\rm mx}^{1,2}$ are the mixing reflections from left to left ($R_{\rm mx}^{1}$) and right to right ($R_{\rm mx}^{2}$).  Eq.(\ref{eq:circuit1}) and Eq.(\ref{eq:circuit2}) define the generalized circuit
theory~\cite{General_Circuit_Theory} so that in the absence of spin-flip scattering C-RMT and generalized circuit
theory are completely equivalent. In fact, the renormalization coefficients of generalized circuit theory~\cite{General_Circuit_Theory} were chosen
such that the calculation of the conductance with RMT and generalized circuit theory fully agree with each other.
We find that the point of view of scattering taken in this letter is fully equivalent to the alternative view in 
term of local current and chemical potential (VF, circuit theory), and one can change from one to the other simply using Eq.(\ref{eq:j}) and Eq.(\ref{eq:mu}). We note that this mapping is quite general and can be extended to include for instance interface spin-flip scattering or superconductivity~\cite{SuperWaintal} .
\begin{figure}
\includegraphics[keepaspectratio,width=0.85\linewidth]{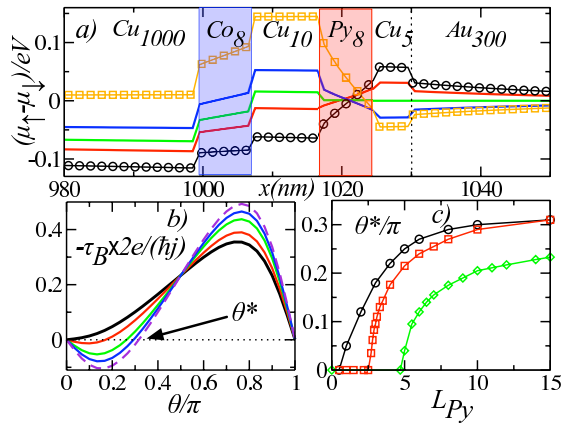}
\caption{\label{fig:c-rmt}
a) Spin accumulation in the middle of a spin valve
$A=Cu_{1000}Co_{8}Cu_{10}Py_{8}Cu_{5}Au_{300}$ (thickness in nm) for different angle
 $\theta=0$ (circles),$\pi/4$, $\pi/2$, $3\pi/4$ and $\pi$ (squares). Symbols stand
for VF calculations while lines correspond to C-RMT. 
b) torque $\tau_B(\theta)$ (per total current $j$) on the $Py$-layer of $A$ 
for various $Py$-thickness $L_{Py}$ from $0.5$nm (thick) to $15$nm (dashed).
c) Stability angle $\theta^*$ as a function
of  $L_{Py}$ for $A$ (circles), $B=Cu_{1000}Co_{8}Cu_{10}Py_{L_{Py}}Cu_{1000}$ with $\delta=0$ (squares)
and $B$ with $\delta_{CoCu}=\delta_{CoPy}=0.25$ (diamonds).}
\end{figure}

{\it Tabulation of CRMT with Valet Fert set of parameters.} As C-RMT and VF are equivalent, 
we can use the huge corpus of experimental data that has been interpreted 
within VF to parametrize C-RMT.
The Valet-Fert resistivities $\rho_{\uparrow(\downarrow)}= 2\rho^* (1\mp \beta)$ and 
spin-diffusion length $l_{\rm sf}$ are in  a one-to-one correspondence with C-RMT:
${1}/{l_{\rm sf}} = 2\sqrt{\Gamma_{\rm sf}}\sqrt{\Gamma_\uparrow + \Gamma_\downarrow}$,
$\beta=(\Gamma_\downarrow - \Gamma_\uparrow)/(\Gamma_\uparrow + \Gamma_\downarrow)$ and
$\rho^*/{\cal R}_{\rm sh}=(\Gamma_\uparrow+\Gamma_\downarrow)/4$. Similarly, within VF, interfaces are described by a set of three parameters, $r^b_{\uparrow(\downarrow)} = 2 r^{b*} (1\mp\gamma)$ and $\delta$ 
which correspond to an effective layer of thickness $d$ which is taken to be infinitely thin while keeping the parameters $\delta=d/l_{\rm sf}$ and $r^{b}_{\sigma}=\rho_{\sigma} d$ constant. Fig.\ref{fig:c-rmt} (a) shows the spin accumulation profile 
of a Co/Py spin valve calculated with  C-RMT. For the parallel and anti-parallel configuration we also show an 
(independent) Valet-Fert calculation using the CNRS-Thales software developed by one of us (HJ). 
Both match perfectly as expected.

{\it Application to spin torque: Wavy or not wavy?} We now turn to an application of C-RMT to the calculation of spin torque in realistic spin valves, taking int account spin-flip scattering both in the bulk and at the interfaces. We consider the valve depicted in Fig.\ref{fig:cartoon} (a) and note $\vec J_A$, $\vec J_N$
and $\vec J_B$ the spin currents just before, in between and after the two magnetic layers $F_A$ and $F_B$. For non collinear magnetization, spin current is
not conserved and the spin torque on $F_A$ and $F_B$ is defined as,
 \begin{eqnarray}
\label{eq:tauA}
\vec\tau_A=\vec J_A -\vec J_N, \ \ \ 
\vec\tau_B=\vec J_N -\vec J_B
\end{eqnarray}
Let us start with a simple geometric construction that allows to get a physical picture for the torque in a rather general way. To do so, we need 
two hypothesis: (i) the mixing 
transmission are small (it is the case for metallic magnetic layers) so that
$\vec J_A$ and $\vec J_B$ are parallel to the magnetization of $F_A$ and $F_B$
respectively. (ii) the system is thin enough for spin-flip scattering to be ignored in the active region so that $\vec\tau_A$ ($\vec\tau_B$) is perpendicular to $\vec J_A$ ($\vec J_B$). The construction goes as follows, see Fig.\ref{fig:cartoon} (b) and (c): first we plot $\vec J_A$ and $\vec J_B$ which make the same angle $\theta$ as the magnetization of the respective magnetic layers. Then,
we note that $\tau_{tot}\equiv\vec\tau_A+\vec\tau_B=\vec J_A-\vec J_B$ does not depend on the unknown
$\vec J_N$ and points from the tip of  $\vec J_B$ to the tip of $\vec J_A$. The construction of the torque is then straightforward: the two vectors $\vec\tau_A$ and $\vec\tau_B$ are chosen such that
they are perpendicular to their respective layer and their sum goes from the tip of  $\vec J_B$ to the tip of $\vec J_A$. This simple construction gives, in particular, the sign of the torque as
a function of the angle $\theta$. We find that when, say, $J_B > J_A$ 
the torque on the layer with the highest polarization ($F_B$) can become wavy~\cite{Barnas_wavy,Boulle_wavy}, 
i.e. instead of favoring the parallel or 
antiparallel configurations, the torque stabilizes (or destabilizes depending of the direction of 
the current) a configuration with a finite angle $\theta^*$.
The critical angle $\theta^*$ where the torque vanishes verifies $|J_A/J_B|=\cos\theta^*$. On the other
hand, at small angle, one has the following developpement $|J_A/J_B|(\theta)= 1 - \eta\theta^2/2 +...$ (Current conservation imposes $J_A=J_B$ at $\theta=0$ and the ratio is an even function of $\theta$) so that ``waviness'' is found when $\eta>1$. As $\eta=0$ for a symmetric structure, it means that a finite asymmetry  is needed to enforce waviness. 

Without spin-flip scattering, 
we find $\eta=[\gamma_Br^*_B-\gamma_Ar^*_A+(\gamma_B-\gamma_A)r^*_Ar^*_B/{\cal R}_{\rm sh}]/
(\gamma_Br^*_B+\gamma_Ar^*_A)$ where the effective parameters $\gamma_A$ and $r^*_A$ include both the interface and bulk properties of layer $A$.
More generally, the crossover between normal and wavy can be discussed by looking at the small angle expression derived by Fert et al (Eq.(5) in ~\cite{ferttorque}, we omit the ballistic corrections).
It can be obtained by relating the spin accumulation in the spacer for $\theta\ll1$ to spin current and spin accumulation at  $\theta=0$, and then applying Eq.(\ref{eq:circuit2}). It reads,
\be
\label{wavy}
\frac{d\tau_B}{d\theta}|_{\theta=0}= -\frac{\hbar}{e}
\left[
\frac{j_\uparrow - j_\downarrow}{4}|_{\theta=0} + 
\frac{\mu_\uparrow - \mu_\downarrow}{2e{\cal R}_{\rm sh}}|_{\theta=0}\right]
\ee
When, for instance, one crosses the assymetry border from $J_A>J_B>0$ to $J_B>J_A>0$, the spin accumulation,
 proportional to the gradient of the spin current, changes from  negative to positive and
the second term in Eq.(\ref{wavy}) becomes neagative and begins to compensate the first one. The crossover 
from normal to wavy occurs when, by a further increase of assymetry, 
the spin accumulation term wins and reverse the sign of $d\tau_B/d\theta|_{\theta=0}$
(see Gmitra and Barna\'s~\cite{Barnas_wavy2} for an extensive discussion of the normal to wavy crossover).

Typical examples of our numerical results are presented in Fig.\ref{fig:c-rmt} for Co/Cu/Py samples
in which the asymmetry comes from the short $l_{\rm sf}$ and large polarization and resistivity
of Py. Starting from a small value of $L_{Py}$, an increase of the assymetry and finally a
crossover to wavy ($\theta^*\ne 0$) can be obtained by increasing $L_{Py}$ as shown in 
Fig.\ref{fig:c-rmt} (b) and (c). By comparing the curves in samples with and without Au on the right of the valve, one sees that the short $l_{\rm sf}$ of Au in a layer close to Py tends to increase the asymmetry and the waviness. On the other hand, interface spin-flip is found to favour a normal spin torque. In the experimental results of Boulle et al~\cite{Boulle_wavy} a wavy behaviour was found for Cu/Co/Cu/Py/Cu/Au structures with equal thicknesses (8 nm) for Co and Py.

We thank A. Brataas, G. Bauer and T. Valet, for fruitful discussions, S. Petit-Watelot for his insights
on the geometrical construction of the torque. This work was supported by EC Contract 
No. IST-033749 ``DynaMax".

\end{document}